\documentclass[twocolumn,reprint,aps,floatfix,dvipdfmx]{revtex4-1}
\usepackage{amsmath,amssymb}
\usepackage{graphicx}
\usepackage{bm}
\usepackage{braket}
\usepackage{physics}
\usepackage{color}
\usepackage{dcolumn}
\usepackage{ulem}

\def\k{\bm k}
\def\q{\bm q}
\def\r{\bm r}
\def\<{\langle}
\def\>{\rangle}
\def\({\left(}
\def\){\right)}

\begin{document}

\title{Emergence of pure spin current in doped excitonic magnets}

\author{Shunsuke Yamamoto$^1$}
\author{Koudai Sugimoto$^2$}
\author{Yukinori Ohta$^1$}
\affiliation{
$^1$Department of Physics, Chiba University, Chiba 263-8522, Japan\\
$^2$Department of Physics, Keio University, Yokohama 223-8522, Japan
}

\date{\today}

\begin{abstract}

An excitonic magnet hosts a condensate of spin-triplet excitons composed of conduction-band 
electrons and valence-band holes, and may be described by the two-orbital Hubbard model.  
When the Hamiltonian has the nearest-neighbor interorbital hopping integrals with $d$-wave 
symmetry and the number of electrons is slightly away from half filling, the $\bm{k}$-space spin 
texture appears in the excitonic phase with a broken time-reversal symmetry.  We then show 
that, applying electric field to this doped excitonic magnet along a particular direction, a pure 
spin current emerges along its orthogonal direction.  We discuss possible experimental realization 
of this type of the pure spin current in actual materials.  
\end{abstract}

\maketitle

\section{Introduction}

In a pure spin current, the flow of electrons with up-spin goes along the opposite direction 
to that with down-spin and the net charge current is absent.  Generation of the pure spin 
current is one of the key issues in spintronics applications.  
Although in the past the spin-Hall effect \cite{Maekawa2013JPSJ, Sinova2015RMP} in 
nonmagnetic materials composed of heavy atoms with a strong spin-orbit coupling is employed 
for this purpose, it has recently been suggested that, even in the case where the spin-orbit 
coupling is negligible, the pure spin current can be induced in antiferromagnetically ordered 
systems, such as noncolinear antiferromagnets \cite{Zelezny2017PRL, Zhang2018NJP, Kimata2019Nature} 
and organic colinear antiferromagnets with a broken glide symmetry \cite{Naka2019NC}.  

In this paper, we will show that the pure spin current can also be generated in a spin-triplet 
excitonic phase (EP), where the electrons in the valence band and holes in the conduction 
band form spin-triplet pairs by attractive Coulomb interaction, and condense into a state 
with quantum coherence at low temperatures.  Kune\v{s} and Geffroy \cite{Kunes2016PRL} 
used a doped two-orbital Hubbard model with cross-hopping integrals, i.e., the nearest-neighbor 
hopping terms between different orbitals, and showed that the $\bm{k}$-space spin texture 
emerges in the spin-triplet EP.  They found that the symmetry of the spin texture depends 
on the symmetry of the cross-hopping integrals; in particular, when the symmetry of the 
cross-hopping terms is $p$-wave, an orbital off-diagonal component of the global spin 
current becomes finite in the EP.  However, unfortunately, the net spin current must be zero 
in an equilibrium state \cite{Geffroy2018PRB, Nishida2019PRB}, just as the Bloch's theorem 
claims the absence of any spontaneous currents \cite{Ohashi1996JPSJ}.

To extract the pure spin current in excitonic magnets, an external field must be applied to the system.
We focus on the case where the cross-hopping integrals have the $d$-wave 
symmetry.  In this case, after the EP transition, the state loses the time-reversal 
symmetry, while the net magnetization is zero \cite{Kunes2016PRL}.  As a result, the pure 
spin current is expected to emerge when an external electric field is applied.  

The rest of this paper is organized as follows.  
In Sec.~\ref{Sec:dK1xfUty}, we introduce the two-orbital Hubbard model with the $d$-wave 
cross-hopping terms defined on the two-dimensional square lattice, and give the definition 
of the excitonic order parameter.  The electric and Hall conductivities obtained from the 
linear response theory are also introduced in this section.  In Sec.~\ref{sec:Wgvau9Yf}, 
we calculate the spin currents generated in the EP by varying the interaction parameters 
and density of electrons.  Possible experimental realization of the pure spin current in 
doped excitonic magnets is also discussed in this section.  A summary of our results is 
given in Sec.~\ref{sec:2XWRnsey}. Appendices are given to discuss details of the mean-field 
analysis of the Hamiltonian and the current-current correlation function used.  

\section{Model and method}
\label{Sec:dK1xfUty}

\subsection{Hamiltonian}

\begin{figure}
\begin{center}
\includegraphics[width=\columnwidth]{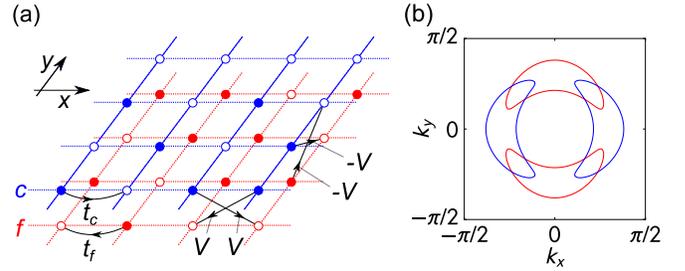}
\end{center}
\caption{
(a) Schematic representation of the kinetic term of the two-orbital Hubbard model 
with $d$-wave cross-hopping integrals [see Eq.~(\ref{eq:H_0}) in the main text].  
The red (blue) circles indicate the $f$ ($c$) orbital.  (b) Calculated $\bm{k}$-space 
spin texture with the $d$-wave symmetry in the spin-triplet EP away from half filling.  
The red (blue) lines denote the Fermi surfaces for up (down) spin.  The interaction 
strengths are set to be $U = 9.5$, $U' = 5.0$, and $J = J' = 0$, and the density 
of electrons is set to be $N = 1.92$.  
}\label{fig:model}
\end{figure}

We consider the two-orbital Hubbard model defined on the two-dimensional square lattice, 
where we include the cross hopping integrals as well as the standard nearest-neighbor 
hopping integrals \cite{Kunes2014PRB1, Kunes2014PRB2, Kunes2016PRL}.  
The Hamiltonian is written as 
\begin{equation}
 \mathcal{H}
	 = \mathcal{H}_0 + \mathcal{H}_\mathrm{int}
\end{equation}
with the kinetic term
\begin{align}
\mathcal{H}_0&=
	\sum_{j, \tau, \sigma}
		\left(t_c c^{\dagger}_{j+\tau,\sigma}c_{j,\sigma}+t_{f} f_{j+\tau,\sigma}^{\dagger}f_{j,\sigma} + \mathrm{H.c.} \right)
\notag \\
	& + \sum_{j,\tau,\sigma}
		\left(V_{1,\tau}c^\dagger_{j+\tau,\sigma}f_{j,\sigma}
		+ V_{2,\tau}f^\dagger_{j+\tau,\sigma}c_{j,\sigma}+\mathrm{H.c.} \right)
\notag \\
	& + \frac{D}{2} \sum_{j,\sigma} \left( n^{c}_{j,\sigma}-n^{f}_{j,\sigma} \right)
\label{eq:H_0}
\end{align}
and the interaction term
\begin{align}
\mathcal{H}_\mathrm{int}
	&= \frac{U}{2} \sum_{i,\sigma} \left(
		f_{i, \sigma}^\dagger  f_{i, \sigma} f_{i, -\sigma}^\dagger f_{i, -\sigma}
		+ c_{i, \sigma}^\dagger  c_{i, \sigma}   c_{i, -\sigma}^\dagger   c_{i, -\sigma}
		\right)
\notag \\
	& + U' \sum_{i, \sigma, \sigma'} 
		f_{i, \sigma}^\dagger  f_{i, \sigma}  c_{i, \sigma '}^\dagger c_{i, \sigma'}
        - J \sum_{i, \sigma, \sigma'} 
		f_{i, \sigma}^\dagger  f_{i, \sigma '}  c_{i, \sigma'}^\dagger  c_{i, \sigma}
\notag \\
	& + \frac{J'}{2} \sum_{i, \sigma} \left(
		f_{i, \sigma}^\dagger    c_{i, \sigma}  f_{i, -\sigma}^\dagger   c_{i, -\sigma}
		+ c_{i, \sigma}^\dagger  f_{i, \sigma}  c_{i, -\sigma}^\dagger   f_{i, -\sigma}
		\right),
\label{eq:H_int}
\end{align}
where $c_{i,\sigma}^{\dagger}$ ($f_{i,\sigma}^{\dag}$) and $c_{i,\sigma}^{}$ ($f_{i,\sigma}^{}$) 
are the creation and annihilation operators of an electron on the conduction-band orbital $c$ 
(valence-band orbital $f$) at site $j$ with spin $\sigma$, and $n^{c}_{j,\sigma}$ ($n^{f}_{j,\sigma}$) 
is the electron number operator on the $c$ ($f$) orbital at site $j$ with spin $\sigma$.  
In Eq.~(\ref{eq:H_0}), $D$ is the on-site energy-level splitting, $t_c$ and $t_f$ are the hopping 
integrals between the same orbitals on the nearest-neighbor sites, $V_{1, \tau}$ and $V_{2, \tau}$ 
are the hopping integrals between the different orbitals on the nearest-neighbor sites 
(which are referred to as the cross-hopping integrals), and $\tau$ denotes the primitive translation vector $\bm{a}_\tau$.  For the cross-hopping integrals with the $d$-wave symmetry, 
we assume $V_{1, x} = V_{2,x} = V_{1,-x} = V_{2,-x} = V$ and $V_{1, y} = V_{2,y} = V_{1,-y} = V_{2,-y} = -V$, 
where $-\tau$ denotes the primitive translation vector of an opposite sign, i.e., 
$\bm{a}_{-\tau} = -\bm{a}_{\tau}$.  The kinetic term is schematically illustrated in 
Fig.~\ref{fig:model}(a).  In Eq.~(\ref{eq:H_int}), $U$, $U'$, $J$, and $J'$ represent the intraorbital 
Coulomb interaction, interorbital Coulomb interaction, Hund's rule coupling, and pair hopping, 
respectively.  

We assume the periodic boundary condition.  Then, the Fourier transformation of 
Eq.~(\ref{eq:H_0}) reads 
\begin{multline}
\mathcal{H}_0 = \sum_{k,\sigma}
\biggl[
	{\varepsilon}_{c}(\bm{k})  {c}_{{\bm k},\sigma}^{\dagger}  {c}_{{\bm k},\sigma}
	+ {\varepsilon}_{f}(\bm{k})  {f}_{{\bm k},\sigma}^{\dagger}  {f}_{{\bm k},\sigma}
\\
	+ \left( {\gamma}(\bm{k}) {c}_{{\bm k},\sigma}^{\dagger}  {f}_{{\bm k},\sigma} + \mathrm{H.c} \right)
\biggr],
\end{multline}
where 
\begin{align}
{\varepsilon}_{c}(\bm{k})
	&= 2t_c\sum_\tau \cos{k_\tau}+\frac{D}{2}, \\
{\varepsilon}_{f}(\bm{k})
	&= 2t_f\sum_\tau \cos{k_\tau}-\frac{D}{2},
\end{align}
and
\begin{equation}
 \gamma (\bm{k}) = 2 V \left( \cos {k_x} - \cos {k_y} \right)
 \label{eq:gamma}
\end{equation}
with $k_\tau={\bm k} \cdot \bm{a}_\tau$.
Hereafter, we assume the hopping integrals as $-t_c = t_f = 1$ (direct gap and 
the unit of energy), which leads to a uniform excitonic order.  We set $D = 6$, 
$V = 0.1$, and the lattice constant to be unity 
($\left| \bm{a}_\tau \right| = 1$) throughout the paper.  

\subsection{Order parameters}

We assume the spin-triplet excitonic order of the spin direction along $z$-axis.
Note that the energy of the spin-singlet excitonic order is strictly equal to that 
of the spin-triplet excitonic order if the Hund's rule coupling is absent and that 
the finite Hund's rule coupling stabilizes the spin-triplet excitonic order 
\cite{Kaneko2014PRB}.  We do not consider the spin-singlet excitonic order here, 
which may be stabilized in the presence of strong electron-phonon coupling terms 
\cite{Kaneko2015PRB}.  

Since the present model has a direct gap, we ignore any excitonic density-wave 
states.  We instead consider the uniform spin-triplet excitonic order, of which the 
order parameter is defined as 
\begin{equation}
\Phi^\mathrm{t}=|\Phi^\mathrm{t}|e^{i\phi} 
=\frac{1}{L^2}\sum_{j, \sigma}\sigma \left\langle {c}_{j, \sigma}^{\dagger} {f}_{j, \sigma} \right\rangle,
\label{eq:spin-triplet-excitonic-order-parameter}
\end{equation}
where $\phi$ is the phase of the complex order parameter, 
$\sigma = 1$ ($-1$) indicates the up (down) spin, and 
$L^2$ is the number of sites in the system.
We apply the mean-field approximation to the Hamiltonian (see Appendix \ref{sec:MFH}) 
and solve the self-consistent equations to obtain the order parameter.  
The representative $d$-wave spin texture in the EP at a hole-doped region 
is illustrated in Fig.~\ref{fig:model}(b).  We find that the Fermi surfaces for each spin 
are anisotropic, implying that the time-reversal symmetry, which exists in the original 
Hamiltonian, is apparently broken in the EP.

In the definition of the spin-triplet excitonic order parameter, we assume that the real and imaginary parts of the order parameter point along the same direction.
However, we should note that there is no restriction to the relative direction between them, except for the case where the pair hopping, which makes the order parameter real \cite{Kaneko2015PRB}, is present.
Since the aim of our study is to discuss the spin current generated in the excitonic phase with $\bm{k}$-space spin texture, we restrict ourselves to considering such a definition, as in Ref. \cite{Nishida2019PRB}, for simplicity.

We also consider the competition between the EP and antiferromagnetic (AFM) 
phase \cite{Kaneko2012PRB}.  The AFM order parameter is defined as 
\begin{equation}
 m_{\ell}
	=\frac{1}{L^2}\sum_{j, \sigma}
		\sigma e^{-i \bm{Q} \cdot \bm{r}_j} \left\langle {\ell}_{j, \sigma}^{\dagger} {\ell}_{j, \sigma} \right\rangle,
\end{equation}
where $\ell$ ($ = c, f$) is an orbital index, $\bm{r}_j$ is the position of site $j$, and 
$\bm{Q} = (\pi, \pi)$ is a checkerboard-type AFM ordering vector.  

Even though the mean-field approximation used here is not sufficient in intermediate or strong coupling regime, it is suitable for obtaining the excitonic ground state with $\bm{k}$-space spin texture.
This is because any symmetry breaking of the system can in principle be described in this approximation,  irrespective of its coupling strength.
For simplicity, we neglect other possible phases such as incommensurate spin-density waves 
and superconductivity here.  
The statistical average $\left\langle \cdots \right\rangle$ is taken 
at absolute zero temperature.  The computations are performed with $L^2 = 400^2$.  

\subsection{Charge and spin currents}

We introduce the external electric field via the Peierls phase.  
The electron operators are then changed as 
\begin{equation}
 \ell_{j, \sigma} \rightarrow e^{i \bm{A} \cdot \r_j} \ell_{j, \sigma},
\end{equation}
where $\bm{A}$ is a vector potential.
The speed of light $c$, reduced Planck constant $\hbar$, 
and elementary charge $e$ are all set to unity.  
We assume a spatially uniform vector potential.
Only the hopping terms in the kinetic term of the Hamiltonian are 
modified by the Peierls substitution.  
The electric current operator may be defined as 
\begin{equation}
\bm{j} = - \left. \frac{\partial \mathcal{H} (\bm{A})}{\partial \bm{A}} \right|_{\bm{A} = 0} = \sum_{\sigma} \bm{j}_{\sigma},
\end{equation}
where
\begin{multline}
 \bm{j}_{\sigma} = i \sum_{j,\tau} \bm{a}_{\tau} \Bigl[
		\left( t_c c^{\dagger}_{j+\tau,\sigma}c_{j,\sigma}+t_{f} f_{j+\tau,\sigma}^{\dagger}f_{j,\sigma} \right)
\\
	+ \left(V_{1,\tau}c^\dagger_{j+\tau,\sigma}f_{j,\sigma}
		+ V_{2,\tau}f^\dagger_{j+\tau,\sigma}c_{j,\sigma} \right) \Bigr] +\mathrm{H.c.}
\end{multline}
indicates the current of the conduction electrons with spin $\sigma$.
If the system shows a real-space spin texture \cite{Loss1990PRL} or the Hamiltonian has spin-orbit interaction terms \cite{Shi2006PRL}, the spin current should be defined as a second-rank pseudotensor given by the flow direction of electron spin and the orientation of the spin.
These effects are not included in our model, and we assume that both the real and imaginary parts of the spin-triplet excitonic order parameter point along $z$-axis.
Hence, we can simply define a spin current as $\bm{j}_{\mathrm{s}} = \bm{j}_{\uparrow} - \bm{j}_{\downarrow}$.
We call the electric current a charge current $\bm{j}_{\mathrm{c}} = \bm{j}_{\uparrow} + \bm{j}_{\downarrow}$.

\subsection{Conductivity}

We carry out the calculation of the spin and charge currents within the linear response theory, where the response function with regard to the applied external field can be obtained from the equilibrium state.
This implies that the external field is assumed to be weak enough, so that the order parameter is not affected. 
The Hamiltonian is expanded with respect to $\bm{A}$ as \cite{Jaklic2000AP}
\begin{equation}
 \mathcal{H} (\bm{A})
	= \mathcal{H} (0) - \bm{j} \cdot \bm{A} - \frac{1}{2} \bm{A} \cdot \underline{u} \bm{A} 
     + \mathcal{O} \( \left| \bm{A} \right|^3 \), 
\end{equation}
where 
 \begin{multline}
 u_{\alpha \beta}
	= \sum_{j,\tau, \sigma} a^{(\alpha)}_{\tau} a^{(\beta)}_{\tau} \Bigl[
		\left( t_c c^{\dagger}_{j+\tau,\sigma} c_{j,\sigma}+t_{f} f_{j+\tau,\sigma}^{\dagger} f_{j,\sigma} \right)
\\
	+ \left(V_{1,\tau}c^\dagger_{j+\tau,\sigma}f_{j,\sigma}
		+ V_{2,\tau}f^\dagger_{j+\tau,\sigma}c_{j,\sigma} \right) \Bigr] +\mathrm{H.c.}
\end{multline}
is the $\alpha\beta$ component of a stress tensor.
We denote $a^{(\alpha)}_{\tau}$ as the $\alpha$-component of $\bm{a}_\tau$.  

The electric field is given by $\bm{E} = - \frac{\partial \bm{A}}{\partial t}$.  
From the linear response theory, the optical conductivity tensor as a function of the 
frequency $\omega$ of the electric field may be given as \cite{Jaklic2000AP}
\begin{equation}
  \sigma_{\alpha\beta} (\omega^+) = \frac{1}{L^2} \frac{i \left[ \chi^{\mathrm{R}}_{\alpha\beta} ( \omega^+) 
  - \< u_{\alpha\beta} \> \right]}{\omega^+} 
\end{equation}
with a retarded current-current correlation function 
\begin{equation}
 \chi^{\mathrm{R}}_{\alpha\beta} (\omega^+)
	= - i \int^{\infty}_{0} \mathrm{d} t \,
		e^{i \omega^+ t} \left\langle  \left[ j^{(\alpha)} (t),  j^{(\beta)} \right] \right\rangle, 
\end{equation}
where $\omega^+ = \omega + i \eta$ (with a positive infinitesimal value $\eta$) and 
$j^{(\alpha)} (t) = e^{i \mathcal{H} t} j^{(\alpha)} e^{-i \mathcal{H} t}$.  
The explicit form of $\chi^{\mathrm{R}}_{\alpha\beta} (\omega^{+})$ in the mean-field approximation is given in Appendix~\ref{sec:I5XZwXob}. 
The Drude weight tensor defined by
\begin{equation}
 D_{\alpha \beta}
	= \frac{\pi}{L^2} \left( \mathrm{Re} \, \chi^{\mathrm{R}}_{\alpha\beta}(0)  - \left\langle u_{\alpha\beta} \right\rangle \right)
\end{equation}
may be separated into contributions from the up-spin and down-spin electrons, i.e.,
\begin{equation}
 D_{\alpha \beta}
 	= \sum_{\sigma} D_{\alpha \beta} (\sigma),
\end{equation}
where $D_{\alpha \beta} (\sigma)$ corresponds to the electric-field response of the spin-$\sigma$ current.
The static conductivity is equal to the real part of the optical conductivity at $\omega = 0$.
In actual materials, there are defects and impurities, which scatter the moving electrons.  
If the scattering rate is approximated to be a constant $\Gamma$, the conductivity tensor 
may be given by \cite{Sugimoto2014PRB} 
\begin{equation}
 \mathrm{Re} \, \sigma_{\alpha \beta} (0) = \frac{D_{\alpha \beta}}{\pi \Gamma}.
 \label{eq:dhUqfm8L}
\end{equation}
When the direction $\alpha$ is parallel (perpendicular) to the direction $\beta$, 
Eq.~(\ref{eq:dhUqfm8L}) represents the electric (Hall) conductivity.  

\section{Results of calculation}
\label{sec:Wgvau9Yf}

\subsection{Pure spin current}

\begin{figure}
\begin{center}
\includegraphics[width=0.8\columnwidth]{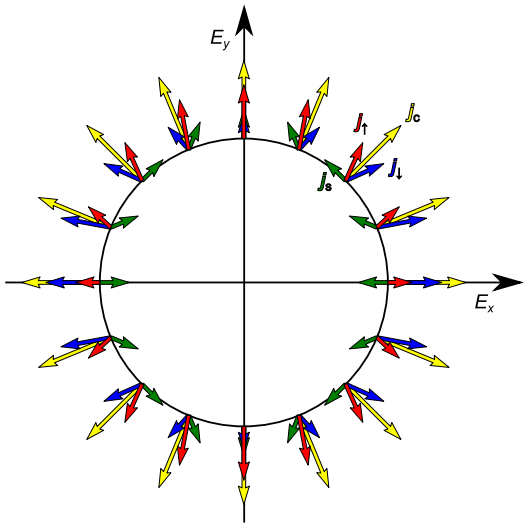}
\end{center}
\caption{
Calculated electric-field-angle dependence of the average values of up-spin 
current $\bm{j}_{\uparrow}$ (red arrows), 
down-spin current $\bm{j}_{\downarrow}$ (blue arrows), 
charge current $\bm{j}_{\mathrm{c}}$ (yellow arrows), 
and spin current $\bm{j}_{\mathrm{s}}$ (green arrows), 
in the case where the $\bm{k}$-space spin texture 
with the $d$-wave symmetry is present.  
The charge current is always parallel to the applied electric field. 
The parameters are set to be the same as used in Fig.~\ref{fig:model}(b).  
}\label{fig2}
\end{figure}

The calculated result for the field-angle dependence of the charge and spin currents in the 
hole-doped ($N=1.92$) spin-triplet EP with the $d$-wave spin texture is illustrated in Fig.~\ref{fig2}.
We find that, reflecting the broken time-reversal symmetry, the spin currents are nonzero in any field angles.  
We also find that, when the electric field is along the diagonal direction of the square lattice, 
the spin current runs along the direction perpendicular to the electric field, while the charge current is parallel to the electric field.  This result clearly indicates that 
the pure spin current is obtained in the doped excitonic magnet when the electric field is applied 
along the diagonal direction of the square lattice.  In other words, the excitonic magnets with the 
broken time-reversal symmetry can host the pure spin current.  

We denote the electric and Hall components of the Drude-weight tensor with spin $\sigma$ 
as $D_{\parallel} (\sigma)$ and $D_{\perp} (\sigma)$, respectively.  In the rest of this section, 
the direction of the electric field is fixed to the $(1,1)$ direction of the square lattice, so that 
we have the identity $D_{\perp} (\uparrow) = - D_{\perp} (\downarrow)$.  
We define the conversion rate of the spin current to the charge current as 
$\beta = \left| D_{\perp} (\uparrow) / D_{\parallel} (\uparrow) \right|$.  
The spin current is present when $\beta \neq 0$, and the value of $\beta$ 
indicates the generation efficiency of the pure spin current with respect to 
the applied electric field.  

\begin{figure}[tb]
\begin{center}
\includegraphics[width=0.9\columnwidth]{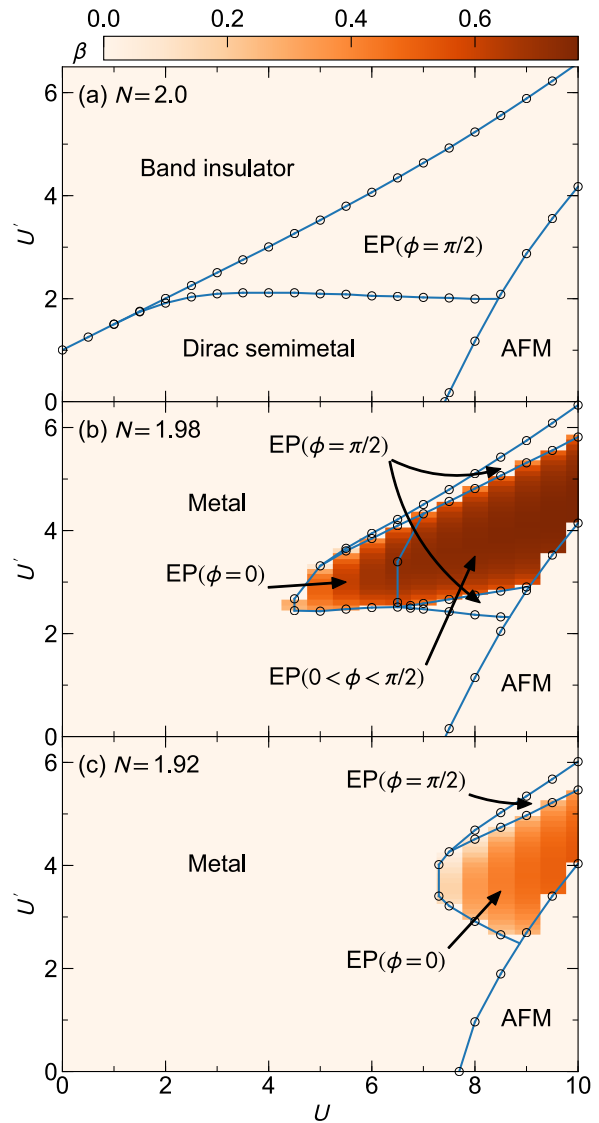}
\end{center}
\caption{
Calculated phase diagrams of the two-orbital Hubbard model at 
(a) $N=2$, (b) $N=1.98$, and (c) $N=1.92$ in the $(U,U')$ plane, 
where we set $J = J' = 0$.
The intensity contour plots indicates the conversion rate $\beta$.
}\label{fig:DM1imOFc}
\end{figure}

\subsection{The case of $J = J' = 0$}

First, let us discuss the case where the Hund's rule coupling and pair hopping term are 
both absent, i.e., $J=J'=0$ in Eq.~(\ref{eq:H_int}).  It is known that the system at half filling ($N=2$) 
without the cross-hopping terms is AFM when $U \gg U'$, while it is band insulating 
when $U' \gg U$, and that the EP emerges between these two phases \cite{Kaneko2012PRB}.

The EP remains even if the cross-hopping terms are introduced.  The phase of the excitonic 
order parameter is fixed to $\phi=\pi/2$ when the cross-hopping integrals have the $d$-wave 
symmetry.  Figure~\ref{fig:DM1imOFc}(a) shows the phase diagram of the two-orbital model 
with the $d$-wave cross-hopping terms at half-filling.  The cross-hopping terms yield the 
Dirac-cone band dispersions in the noninteracting bands \cite{Young2015PRL}, so that the 
four different phases appear in the $U$-$U'$ phase diagram at half-filling; i.e., 
the band insulating phase, normal semimetallic phase, EP with $\phi = \pi / 2$, and the AFM phase.  

As holes are slightly doped into the system, the new EP with $0\leq\phi<\pi/2$ appears.  
Figure~\ref{fig:DM1imOFc}(b) shows the phase diagram at $N=1.98$.  We find that upon doping 
the $\bm{k}$-space spin texture immediately appears in the Fermi surface, which leads to 
the emergence of the spin current.  Note that the k-space spin texture (or the spin current) is originated from the cooperation of the EP transition and cross-hopping integrals and that the spin current remains even when $\phi = 0$ in the hole-doped region.  However, when $\phi = \pi / 2$, the $\bm{k}$-space spin texture disappears in the Fermi surface [9], so that the spin current is no longer observed.  Figure 3(c) shows the phase diagram at $N = 1.92$.  As the doping rate increases, the EP with $0< \phi < \pi / 2$ disappears.
\begin{figure}[tb]
\begin{center}
\includegraphics[width=0.9\columnwidth]{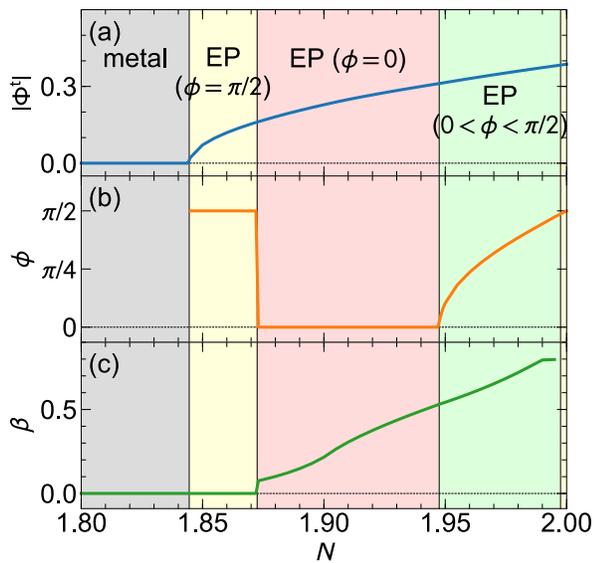}
\end{center}
\caption{
(a) Calculated magnitude of the excitonic order parameter $|\Phi^\textrm{t}|$, 
(b) its phase $\phi$, and (c) conversion rate $\beta$ as a function of the electron 
density $N$.  Interaction strengths are set to $U = 9.5$, $U' = 5.0$, and $J = J' = 0$.  
}\label{fig;MkG3OWEk}
\end{figure}

The excitonic order parameter and conversion rate as a function of the electron 
density at $U = 9.5$ and $U' = 5.0$ are illustrated in Fig.~\ref{fig;MkG3OWEk}.
Both the order parameter and conversion rate monotonically decrease as 
the density of doped holes increases.  We find that, when the $\bm{k}$-space 
spin texture disappears (or equivalently $\beta = 0$), the EP with $\phi = \pi / 2$ 
is stabilized.  In the over-doped region, the EP completely vanishes, and the system 
becomes normal metallic.  

\subsection{The case of finite $J$ and $J'$}

Next, to be more realistic, let us take into account the Hund's rule coupling 
and pair-hopping terms.  We assume the relations $U'=U-2J$ and $J=J'$ 
in this subsection.  It is known that the pair-hopping terms force the phase 
of the excitonic order parameter to be zero \cite{Kaneko2015PRB}.  

\begin{figure}[tb]
\begin{center}
\includegraphics[width=\columnwidth]{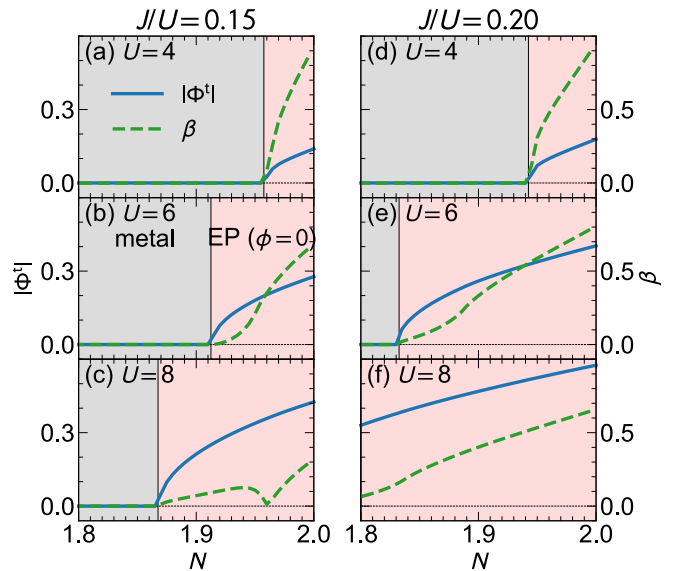}
\end{center}
\caption{Calculated excitonic order parameters $|\Phi^\mathrm{t}|$ (blue solid lines) and 
conversion rate $\beta$ (green dashed lines) as a function of the electron density $N$.  
We assume $J/U = 0.15$ in (a)-(c) and $J/U=0.20$ in (d)-(f), 
and $U=4$ in (a) and (d), $U=6$ in (b) and (e), and $U=8$ in (c) and (f).  
}\label{fig:RSTIbKto}
\end{figure}

Figure \ref{fig:RSTIbKto} shows the calculated results for the excitonic order parameter 
and conversion rate as a function of the electron density.  We find that the spin current 
appears even when finite $J$ and $J'$ are present.  However, the conversion rate is not 
proportional to the excitonic order parameter because the rate is not determined by 
the magnitude of the order but is determined by the degree of the spin splitting in 
the $\bm{k}$-space spin texture.  We note that the AFM phase is not stabilized in 
the parameter region shown in Fig.~\ref{fig:RSTIbKto}.  

\subsection{Possible realization of the pure spin current}

Finally, let us discuss possible experimental realization of the spin current in actual 
materials.  A number of transition-metal oxides have been regarded as candidates for 
the excitonic magnets.  For example, some kinds of cobalt oxides with the $d^6$ electron 
configuration were suggested to be the spin-triplet excitonic magnets \cite{Kunes2014PRL, 
Yamaguchi2017JPSJ, Yamaguchi2018PB, Moyoshi2018PRB, Tomiyasu2018AQT}, 
where the electrons in the $e_g$ orbitals and holes in the $t_{2g}$ orbitals form excitonic 
pairs.  Also, Ca$_2$RuO$_4$ with the $d^4$ electron configuration was suggested 
to be an excitonic magnet \cite{Khaliullin2013PRL, Akbari2014PRB, Jain2017NP}.  
However, unfortunately, no experimental observations of the spin texture have so far 
been reported in these materials.  

\begin{figure}[tb]
\begin{center}
\includegraphics[width=0.7\columnwidth]{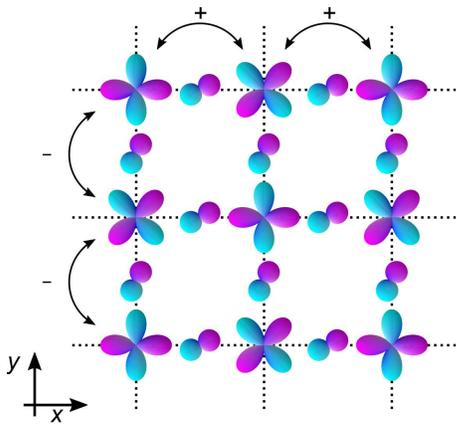}
\end{center}
\caption{Schematic picture of the orbital pattern leading to the $\bm{k}$-space spin 
texture with the $d$-wave symmetry in the EP.  Note that the tilting of the ligand $p$ 
orbitals makes the cross-hopping integrals be of the $d$-wave symmetry.  
}\label{fig:cross}
\end{figure}

In order to realize excitonic magnets with $\bm{k}$-space spin texture of the $d$-wave 
symmetry, we may focus on transition-metal oxides with the layered perovskite structure.  
Figure~\ref{fig:cross} illustrates an orbital pattern, which leads to the cross-hopping integrals 
of the $d$-wave symmetry.  To make the direct-gap band dispersions in the square lattice, 
we may assume that the one band comes from the $d_{xy}$ orbitals and the other comes 
from the $d_{x^2 - y^2}$ (or $d_{3z^2-r^2}$) orbitals.  This is because the nearest-neighbor 
hopping integrals between the $d_{xy}$ orbitals are positive, while those between the 
$d_{x^2 - y^2}$ orbitals are negative \cite{Slater1954PR}.  Note that the cross-hopping 
integrals between the $d_{xy}$ and $d_{x^2 - y^2}$ orbitals via the tilted $p$ orbitals on 
ligand oxygen ions may have the $d$-wave symmetry (see Fig.~\ref{fig:cross}).  Thus, 
we may suggest a possible experimental realization of the pure spin current in such 
transition-metal oxides with the layered perovskite structure.  

\section{Conclusions}
\label{sec:2XWRnsey}

We have studied the pure spin current in the doped excitonic magnets, which is generated 
by applying the external electric field.  The two-orbital Hubbard model defined on the 
two-dimensional square lattice with the cross-hopping integrals of the $d$-wave symmetry 
was solved in the mean-field approximation.  We thus found that the $\bm{k}$-space spin 
texture emerges in the spin-triplet EP when the holes are slightly doped.  Since the spin 
texture breaks the time-reversal symmetry, we found that the pure spin current is generated 
along the orthogonal direction of the charge current when the electric field is applied parallel 
to the diagonal direction of the square lattice.  

We also obtained the phase diagrams of the two-orbital Hubbard model by varying the 
intra- and inter-orbital interactions.  We found that the pure spin current emerges in 
slightly hole-doped EP, which is however suppressed by the excess hole doping.  When 
the phase of the excitonic order parameter becomes $\phi = \pi / 2$, the spin texture 
disappears and consequently the spin current vanishes.  We also investigated the case 
where the Hund's rule coupling and pair-hopping terms are present and confirmed that 
the pure spin current can exist in such realistic cases as well.  We suggested that some 
kinds of $3d$ transition-metal oxides with the layered perovskite structure may host 
the pure spin current driven by the excitonic order, of which the experimental studies 
are desired.  

We want to emphasize that the spin current discussed in this paper is induced purely 
from the effects of electron correlations.  Thus, we found a route to generate the 
pure spin current in correlated electron systems without the spin-orbit coupling.  

\section*{Acknowledgments}
We thank S. Miyakoshi, H. Nishida, and T. Yamaguchi for enlightening discussions.
This work was supported in part by Grants-in-Aid for Scientific Research from 
JSPS (Projects No.~JP17K05530 and No.~JP19K14644) and by Keio University 
Academic Development Funds for Individual Research.

\appendix
\section{Mean-field Hamiltonian}
\label{sec:MFH}

First, let us introduce the expectation value defined as
\begin{equation}
 n^{\ell \ell'}_{\bm{q}} = \sum_{\sigma}  n^{\ell \ell'}_{\bm{q}, \sigma}
	= \frac{1}{L^2} \sum_{\k, \sigma} \left\< \ell_{\k, \sigma}^\dagger \ell'_{\k + \bm{q}, \sigma} \right\>,
\end{equation}
and denote $\bar{c} = f$ and $\bar{f} = c$.  Thereby, $n^{cc}_{0}$ and $n^{ff}_{0}$ 
represent the numbers of electrons on the $c$ and $f$ orbitals per site, respectively.  

The mean-field Hamiltonian of the two-orbital Hubbard model may then be obtained as 
\begin{equation}
 \mathcal{H}^{\rm MF}
	= \mathcal{H}_{0} + \mathcal{H}_{\mathrm{int}}^{\rm MF}, 
\label{eq:Vusk2c7n}
\end{equation}
where
\begin{multline}
 \mathcal{H}_{\mathrm{int}}^{\rm MF}
	= \sum_{\bm{k}, \bm{q}, \sigma}
\biggl[
	 \tilde{n}^{cc}_{\q, \sigma}  {c}_{{\bm k} + \bm{q},\sigma}^{\dagger}  {c}_{{\bm k},\sigma}
	+  \tilde{n}^{ff}_{\q, \sigma}  {f}_{{\bm k} + \bm{q},\sigma}^{\dagger}  {f}_{{\bm k},\sigma}
\\
	+ \tilde{n}^{fc}_{\q, \sigma} {c}_{{\bm k} + \bm{q},\sigma}^{\dagger} {f}_{{\bm k},\sigma}
	+ \tilde{n}^{cf}_{\q, \sigma} {f}_{{\bm k} + \bm{q},\sigma}^{\dagger} {c}_{{\bm k},\sigma}
\biggr] + L^2 \epsilon_0
\end{multline}
with
\begin{align}
 \tilde{n}^{\ell \ell}_{\q, \sigma}
 	&= U n^{\ell \ell}_{\q, -\sigma} + U' n^{\bar{\ell} \bar{\ell}}_{\q} - J n^{\bar{\ell} \bar{\ell}}_{\q, \sigma},
\\
 \tilde{n}^{\ell \bar{\ell}}_{\q, \sigma}
 	&= - U' n_{\q, \sigma}^{\ell \bar{\ell}}
		+ J \Phi_{\q}^{\ell \bar{\ell}} +  J' n^{\bar{\ell} \ell}_{\q, -\sigma},
\end{align}
and
\begin{align}
 \epsilon_0
 	=& - \frac{U}{2} \sum_{\sigma, \q} \left[ n^{ff}_{\q, -\sigma} \( n^{ff}_{\q, \sigma} \)^*
		+ n^{cc}_{\q, -\sigma} \( n^{cc}_{\q, \sigma} \)^* \right]
\notag \\
		&- U' \sum_{\q}
		\left[  \( n^{ff}_{\q} \)^* n^{cc}_{\q}
		- \sum_\sigma \( n_{\q, \sigma}^{cf} \)^* n_{\q, \sigma}^{cf} \right]
\notag \\
		&+  J \sum_{\q}
		\left[ \sum_\sigma \( n^{ff}_{\q, \sigma} \)^* n^{cc}_{\q, \sigma}
		- \( n_{\q}^{cf} \)^* n_{\q}^{cf} \right]
\notag \\
		&- \frac{J'}{2} \sum_{\sigma, \q}
 	\left[ \( n^{cf}_{\q, \sigma} \)^* n^{fc}_{\q, -\sigma}
 		+ \( n^{fc}_{\q, \sigma} \)^* n^{cf}_{\q, -\sigma} \right].
\end{align}

\subsection{Spin-triplet excitonic order}

The order parameter for the uniform spin-triplet excitonic condensation may be written as
\begin{equation}
\Phi^\mathrm{t} 
	=\sum_{\sigma}\sigma n^{cf}_{\bm{q}=0, \sigma}, 
\end{equation}
where we assume that the terms with $\bm{q} \neq 0$ vanish in the uniform EP.  
Diagonalizing the $2 \times 2$ mean-field Hamiltonian matrix, we obtain a quasiparticle 
operator $\alpha_{\k, \sigma,  \epsilon}^\dagger$ for the band $\epsilon$, which 
creates a quasiparticle with energy $E^{\epsilon}_{\k, \sigma}$.  
The quasiparticle operators satisfy 
$\mathcal{H}^{\rm{MF}} = \sum_{\k, \sigma} \sum_{\epsilon} 
E^{\epsilon}_{\k, \sigma} \alpha^\dagger_{\bm{k}, \sigma, \epsilon} \alpha_{\bm{k}, \sigma, \epsilon} 
+ L^2 \epsilon_0$ and 
$\ell_{\k, \sigma} = \sum_{\mu} \psi_{\ell; \epsilon} (\k, \sigma)  \alpha_{\bm{k}, \sigma, \epsilon} $.  
The chemical potential $\mu$ is determined from the equation 
\begin{equation}
N=\frac{1}{L^2}\sum_{{\bm k},\sigma} \left[ f(E_{\bm{k},\sigma}^{+})+f(E_{\bm{k},\sigma}^{-}) \right] 
\end{equation}
and the order parameter is obtained by solving the self-consistent equation 
\begin{equation}
\Phi^\mathrm{t} = \frac{1}{L^2}\sum_{{\bm k},\sigma, \epsilon} \psi^*_{c; \epsilon} 
\psi_{f; \epsilon} f(E_{\bm{k},\sigma}^{\epsilon}), 
\end{equation}
where we define the Fermi distribution function as 
$f(E_{\bm{k}, \sigma}^{\pm}) =\langle {\hat{\alpha}}_{\bm{k},\sigma,\pm}^{\dagger} {\hat{\alpha}}_{\bm{k}, \sigma, \pm} \rangle 
=1 / \left(1+e^{\beta \left( E_{\bm{k}, \sigma}^{\pm} - \mu \right)} \right)$ 
with the reciprocal temperature $\beta$.  
Details of the mean-field calculation of the two-orbital Hubbard model are given in 
Ref.~\cite{Nishida2019PRB}.  

\subsection{AFM order}

The order parameter for the AFM state may be written as 
\begin{equation}
 m_{\ell}	 =\sum_{\sigma} \sigma n^{\ell \ell}_{\bm{Q}, \sigma}
\end{equation}
with $\bm{Q} = (\pi, \pi)$.  The $\bm{k}$-summation over the Brillouin zone was 
rewritten as $\sum_{\bm{k}} \rightarrow  \sum_{\bm{k}_0} \sum_{m = 0, 1}$ with 
$\bm{k} \rightarrow \bm{k}_0 + m \bm{Q}$, where the sum of $\bm{k}_0$ was taken 
over the reduced AFM Brillouin zone.  Diagonalizing the $4 \times 4$ 
mean-field Hamiltonian matrix for each wave vector and spin, we obtain a quasiparticle 
operator $\alpha_{\k_0, \sigma,  \epsilon}^\dagger$ for the band $\epsilon$, which 
creates a quasiparticle with energy $E^{\epsilon}_{\k_0, \sigma}$.  
The quasiparticle operators satisfy 
$\mathcal{H}^{\rm{MF}} = \sum_{\k_0, \sigma} \sum_{\epsilon} 
E^{\epsilon}_{\k_0, \sigma} \alpha^\dagger_{\bm{k}_0, \sigma, \epsilon} \alpha_{\bm{k}_0, \sigma, \epsilon} 
+ L^2 \epsilon_0$ and 
$\ell_{\k_0 + m \bm{Q}, \sigma} = \sum_{\mu} \psi_{\ell, m; \epsilon} (\k_0, \sigma)  \alpha_{\bm{k}_0, \sigma, \epsilon}$.
The chemical potential is determined from the equation 
$N=(1/L^2)\sum_{{\bm k}_0,\sigma, \epsilon} f(E_{\bm{k}_0,\sigma}^{\epsilon})$ 
and the order parameter is obtained by solving the self-consistent equations.  

\section{Current-current correlation function}
\label{sec:I5XZwXob}

Using the results of the mean-field approximation given in Appendix \ref{sec:MFH}, we obtain 
the current operator as 
\begin{equation}
 \bm{j}
	= \sum_{\k_0, \sigma} \sum_{\epsilon, \epsilon'} \bm{J}_{\epsilon, \epsilon'} (\k_0, \sigma)
		\alpha^\dagger_{\k_0, \sigma, \epsilon}  \alpha_{\k_0, \sigma, \epsilon'}
\end{equation}
with
\begin{multline}
 \bm{J}_{\epsilon, \epsilon'} (\k_0, \sigma)
	= i \sum_m  \sum_{\tau}  \sum_{\ell,\ell'}
		\bm{a}_{\tau} t_{\ell \ell'} (\tau) \\
		\times e^{ - i (\k_0 + m \bm{Q}) \cdot \bm{a}_{\tau}}
			\psi^*_{\ell, m ; \epsilon} (\k_0, \sigma) \psi_{\ell', m ; \epsilon'} (\k_0, \sigma), 
\end{multline}
where we define $t_{\ell \ell}  (\tau) = t_{\ell}$, $t_{cf} (\tau) = V_{1,\tau}$ and $t_{fc} (\tau) = V_{2, \tau}$.  

Using the matrix elements of the current operator, we obtain the current-current correlation function as 
\begin{multline}
\chi^{\mathrm{R}}_{\alpha\beta}(\omega^+) 
	=  \sum_{\bm{k}_0, \sigma} \sum_{\epsilon, \epsilon'}
		\frac{f(E^{\epsilon}_{\bm{k}_0, \sigma} ) - f(E^{\epsilon'}_{\k_0, \sigma} )}{(E^{\epsilon}_{\k_0, \sigma} - E^{\epsilon'}_{\k_0, \sigma}) + \omega^+}
\\
		\times J^{(\alpha)}_{\epsilon, \epsilon'} (\k_0, \sigma) J^{(\beta)}_{\epsilon', \epsilon} (\k_0, \sigma), 
\end{multline}
where the sum of $\bm{k}_0$ is taken over the original Brillouin zone and $m=0$ for the uniform excitonic 
state, but it is taken over the reduced Brillouin zone and $m=0, 1$ for the AFM state.

\end{document}